\documentclass[final,3p,times,twocolumn]{elsarticle}
\biboptions{square,sort&compress}

\begin{document}

\begin{frontmatter}



\title{Vortex-antivortex annihilation in mesoscopic superconductors with a central pinning center}

\author[1] {R.\ Zadorosny}
\author[1] {E.\ C.\ S.\ Duarte}
\author[2,3] {E.\ Sardella}
\author[4] {W.\ A.\ Ortiz}

\address[1] {Grupo de Desenvolvimento e Aplica\c{c}\~{o}es de Materiais, Faculdade de Engenharia, \textit{Univ Estadual Paulista - Unesp}, Departamento de Física e Química, Caixa Postal 31, 15385-000, Ilha Solteira, SP, Brazil}

\address[2] {Faculdade de Ci\^{e}ncias,  \textit{Univ Estadual Paulista - Unesp}, Departamento de F\'{i}sica, Caixa Postal 473, 17033-360, Bauru, SP, Brazil}

\address[3] {\textit{Univ Estadual Paulista - Unesp}, Instituto de Pesquisas Metereol\'ogicas - IPMet, CEP 17048-699, Bauru, SP, Brazil}

\address[4] {Departamento de F\'{i}sica, \textit{ Universidade Federal de S\~{a}o Carlos - UFSCar}, Caixa Postal 676, 13565-905, S\~{a}o Carlos, SP, Brazil}

\begin{abstract}

In this work we solved the time-dependent Ginzburg-Landau equations, TDGL, to simulate two superconducting systems with different lateral sizes and with an antidot inserted in the center. Then, by cycling the external magnetic field, the creation and annihilation dynamics of a vortex-antivortex pair was studied as well as the range of temperatures for which such processes could occur. We verified that in the annihilation process both vortex and antivortex acquire an elongated format while an accelerated motion takes place.

\end{abstract}


\end{frontmatter}


%%\linenumbers

\section{Introduction}
\label{Intro}

  A variety of exotic behaviors which arise from confinement effects in superconducting materials have been extensively studied both for type I \cite{berdiyorov,muller} and type II \cite{schweigert,melnikov,mertelj,baelus,geim1,chibotaru,moshchalkov,sardella2,pascolati,benxu,
cren,golubovic,milosevic,yu,geim3,buzdin,palacios,zad2,misko2,zad1,connolly} superconductors. Particularly, the penetration and propagation of vortices in superconducting materials are issues of great interest both from the theoretical point of view and in applications of these materials. In general, a vortex interacts with the surface of the sample and, after surpassing it; with other vortices, which can already have penetrated the sample; as well as with defects that might be present. Defects, in general, act as pinning centers avoiding the dissipative motion of vortices and, as a consequence, the critical current density, $J_c$, and the upper critical field, $H_{c2}$, of the superconductor can be increased.

  In a previous work, Sardella \textit{et al.}, \cite{sardella4} used the time-dependent Ginzburg-Landau, TDGL, equations to study the annihilation process of a vortex-antivortex pair, V-AV, in a square mesoscopic system with a centered square antidot, AD. As described there, the average velocity of the vortex presents distinct values for different processes, i.e., while entering, its velocity is of the order of $10^3 m/s$ whereas in the annihilation motion the average velocity is of the order of $10^5 m/s$. In the present work we used the TDGL equations to study, in two systems, the temperature range for which a V-AV pair should occur in the superconductor region. Such systems are infinitely long cylinders of square cross section. A square AD with lateral size of $2\xi(0)$ was inserted in the center of the system. The creation of a V-AV pair was done by cycling the magnetic field which was applied parallel to the axis of the cylinder. The field was increased until at least one vortex was trapped in the AD, and then decreased and reversed. In such process, a vortex is trapped in the AD and, upon decrease of the field, one of the following three events could occur, depending on the value of the temperature and the lateral size of the sample: (i) an antivortex penetrates the sample and the annihilation occurs in the AD; (ii) the vortex leaves the AD and the annihilation occurs in the superconducting region, and (iii) the vortex is untied and then leaves the sample. In this work we have studied the conditions under which event (ii) can occur, and analyze the dynamics of the annihilation process. It is interesting to note that, even when the effective superconducting region has the same size of the vortex core, i.e, $d=2\xi(0)$, a V-AV pair is formed and an annihilation process takes place.

  The outline of this work is as follows. First, in section \ref{TDGL}, we provide an overview of the theoretical formalism used to run the simulations. Next, in section \ref{results} we discuss some results obtained for two systems with lateral sizes of $L=6\xi(0)$ and $L=12\xi(0)$ and in section \ref{theend} we present our conclusions.

\section{Theoretical Formalism}
\label{TDGL}

  The superconducting state is described by the Ginzburg-Landau theory through a complex order parameter $\psi$ for which $|\psi|^2$ represents the density of Cooper pairs. In regions where $|\psi|^2$ is small, the superconductivity is suppressed. At the core of a vortex, $|\psi|^2=0$, whereas the local magnetic field $h$ is maximum. To determine $h$ and $\psi$, the time dependent Ginzburg-Landau equations, TDGL, were used:

\begin{eqnarray}
\left ( \frac{\partial}{\partial t} + i\Phi\right ) \psi& = & -\left
(-i\mbox{\boldmath
$\nabla$}-{\bf A} \right )^2\psi\nonumber \\
& & +(1-T)\psi(1-|\psi|^2)\;,\nonumber \\
\beta\left ( \frac{\partial{\bf A}}{\partial t}+\mbox{\boldmath
$\nabla$}\Phi \right ) & = & {\bf J}_s-\kappa^2\mbox{\boldmath
$\nabla$}\times{\bf h}\;,\label{tdgleq}
\end{eqnarray}
where the supercurrent density is given by
\begin{equation}
{\bf J}_s=(1-T)\Re\left [ \psi^{*}\left ( -i\mbox{\boldmath
$\nabla$}-{\bf A} \right )\psi \right ]\;, \label{densityCurrent}
\end{equation}
${\bf A}$ is the vector
potential which is related to the local magnetic field as
${\bf h}=\mbox{\boldmath $\nabla$}\times{\bf A}$, and $\Phi$ is the scalar potential.
As we have worked with the normalized TDGL equations, the distances are measured in units of the coherence length
at zero temperature $\xi(0)$; the magnetic field is in units of the zero temperature upper critical field
$H_{c2}(0)$; the temperature $T$ is in units of the critical
temperature $T_c$; the time is in units of the characteristic
time $t_0=\pi\hbar/8k_BT_c$; $\kappa$ is the Ginzburg-Landau parameter;
$\beta$ is the relaxation time of $\bf A$, related
to the electrical conductivity. We have adopted, for the upper critical field, a linear dependence with respect to temperature,
i.e., $H_{c2}(T)=H_{c2}(0)(1-T)$. For
small size superconductors this is also valid for temperatures well below $T_c$,
despite the microscopic derivation of the TDGL equations being valid
only for $T$ very close to $T_c$ \cite{geurts,geurts2}. Notice that the TDGL equations
\cite{gropp} are gauge invariant under the transformations $\psi^{\prime}=\psi e^{i\chi}$,
${\bf A}^{\prime}={\bf A}+\mbox{\boldmath $\nabla$}\chi$,
$\Phi^{\prime}=\Phi-\partial\chi/\partial t$.
Thus, in particular, we choose the zero-scalar potential gauge, that is,
$\Phi=0$ at all times and positions.

\section{Results and Discussion}
\label{results}

\begin{figure}
\includegraphics[width=0.9\columnwidth,height=0.7\linewidth]{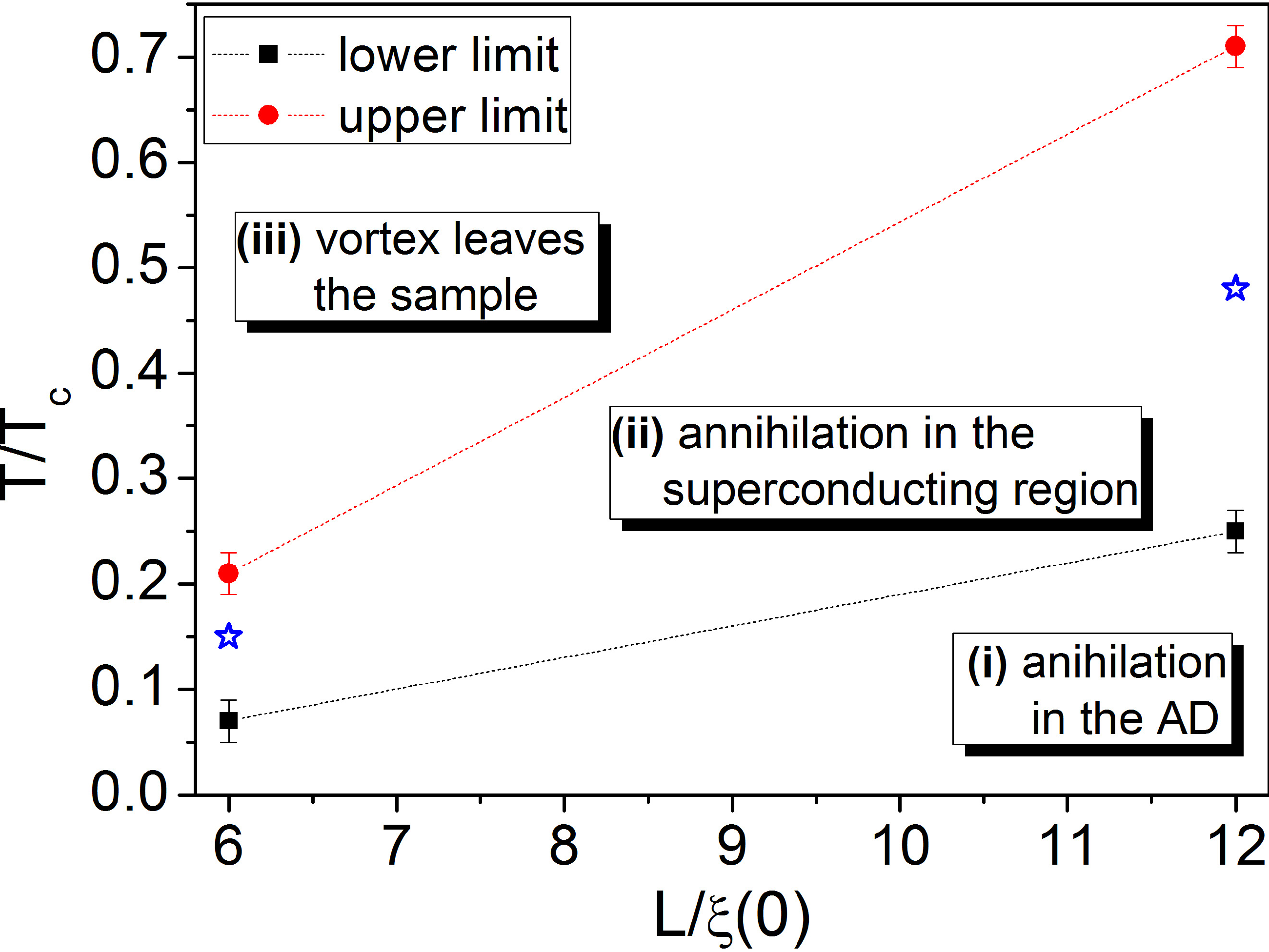}
\caption{(Color online) Limits for the dynamics of vortex-antivortex annihilation in the superconducting region. The dashed lines are only guides for the eyes. The star symbols indicate the systems for which the annihilation dynamics was analyzed.}
\label{fig1}
\end{figure}

 The simulations were carried out by using $\beta=1$ and $\kappa=5$ for two square systems with lateral sizes $L=6\xi(0)$ and $L=12\xi(0)$ where a square AD of side $L=2\xi(0)$ was inserted in the center of the system. The temperature was varied in steps $\Delta T=0.2 T_c$ and the external field in $\Delta H=10^{-3}H_{c2}(0)$. Thus, we found the range of temperatures for which the annihilation of a V-AV pair occurs in the superconducting region. As described in section ~\ref{Intro} the creation of a V-AV pair was done by cycling the applied magnetic field. Thus, one of the three processes described in section \ref{Intro} could occur. Fig.~\ref{fig1} shows the points which delimitate the regions of the tree distinct behaviors, i.e., below the lower points, situation (i) takes place; above the higher points one has the situation (iii); and between such points, option (ii) takes place.

\begin{figure}
\includegraphics[width=0.9\columnwidth,height=1.3\linewidth]{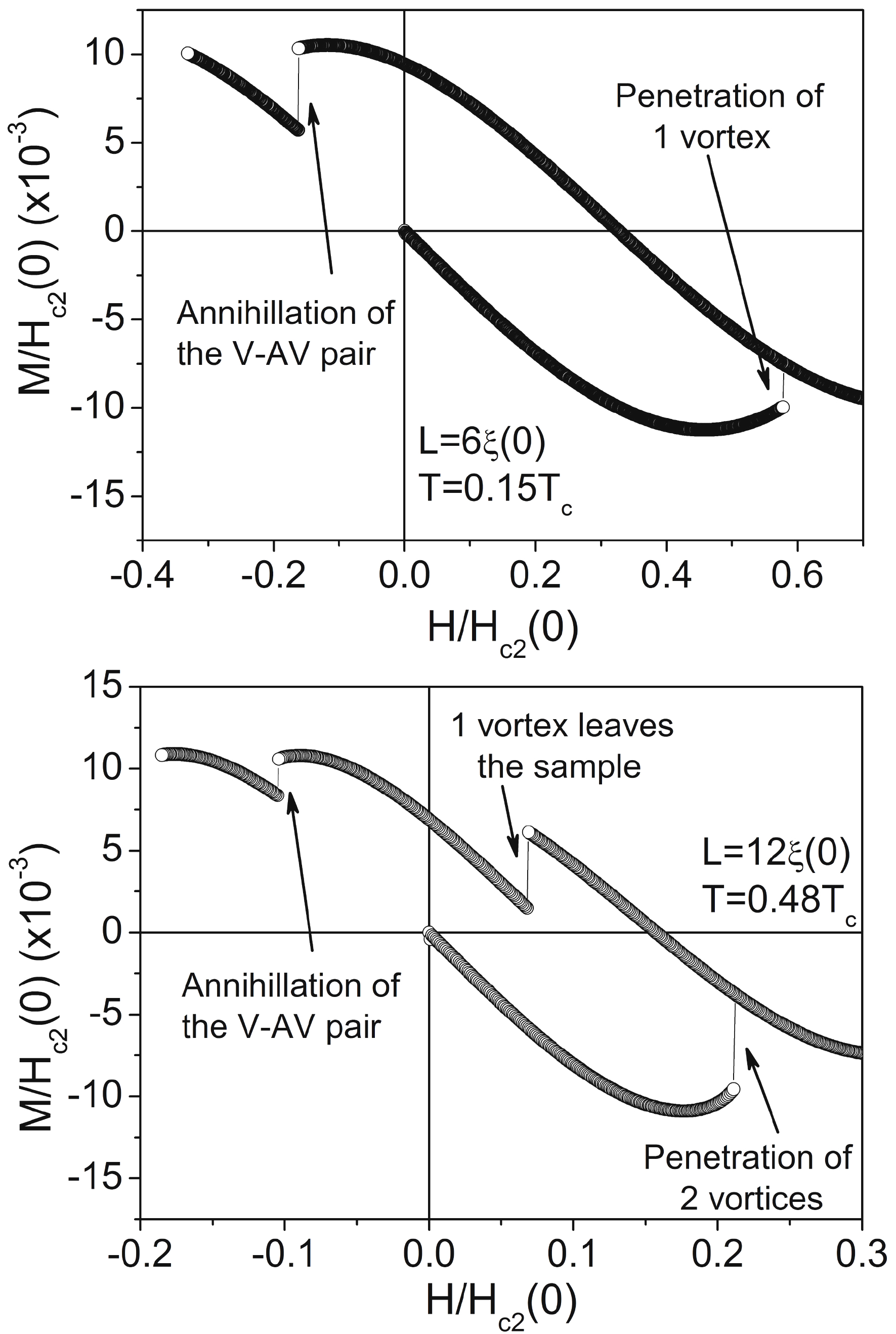}
\caption{Magnetization versus applied magnetic field, both normalized by $H_{c2}$, of the simulated systems $L=6\xi(0)$ and $L=12\xi(0)$. In the larger system, two vortices are nucleated in the first penetration. Then, before the annihilation of the V-AV pair, one vortex is untrapped and leaves the sample.}
\label{fig2}
\end{figure}

 Fig.~\ref{fig2} shows a magnetization versus applied magnetic field curve, both normalized by $H_{c2}(0)$, for the simulated systems. Note that, as the superconducting region of the sample with $L=12\xi(0)$ is greater than that for the sample with $L=6\xi(0)$, two vortices nucleate into the sample in the first penetration, being trapped by the AD. When the field is reversed, firstly one vortex is untied from the AD and then leaves the sample. In the sequence, for negative values of the applied field, another vortex leaves the AD and a V-AV pair is created with the penetrating antivortex. Thus, the annihilation of such pair occurs in the superconducting region.

\begin{figure}
\includegraphics[width=0.8\columnwidth,height=0.6\linewidth]{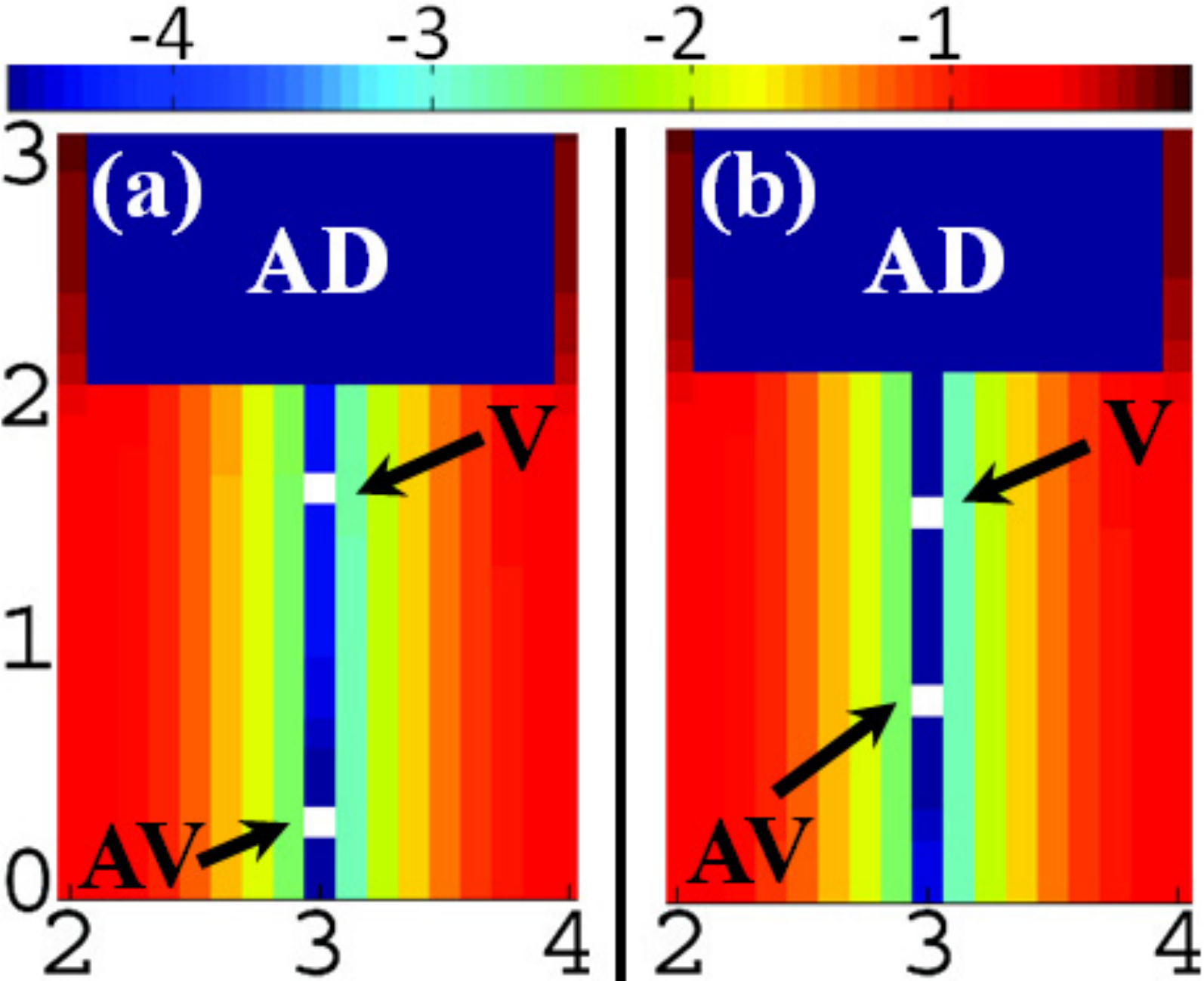}
\caption{(Color online)Images of the logarithm of $\psi$ for the smaller sample.  The white spots indicated the vortex (near the AD) and the AV (near the border of the sample) core, where $\psi=0$. Note the channel (dark) formed by the extension of the V and the AV. The time evolution of the system follows from (a) to (d).}
\label{fig3}
\end{figure}

 Another intriguing behavior is the formation of a V-AV pair in the smaller sample. Such sample has an effective superconducting region with width $d=2\xi(0)$, i.e., the size of the vortex core diameter. However, the presence of the AD distorts the vortex/antivortex profile, i.e., these entities acquired an elongated format and, as the effective superconducting region is very small, such elongations, where the order parameter has small values, overlaps creating a channel between the region outside the system and the AD. In this channel, the order parameter is very small and the cores of the vortex and of the antivortex, with $|\psi|^2=0$, move toward each other until full annihilation. Figure~\ref{fig3} shows such process where, for better visualization, we have plotted the logarithm of $\psi$. The white spots locate the vortex (near the AD) and the antivortex (near the border of the sample).

\begin{figure}
\includegraphics[width=0.9\columnwidth,height=0.8\linewidth]{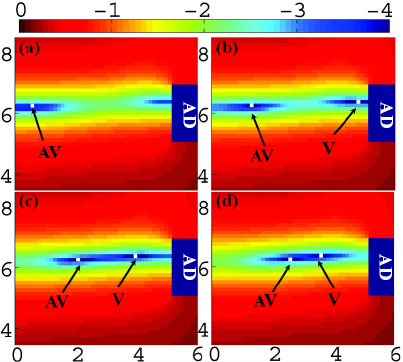}
\caption{(Color online) Images of the logarithm of $\psi$ for the lager sample. The V and the AV are extended, however no channel is formed. The white spots indicate the vortex (near the AD) and the AV (near the border of the sample) core, where $\psi=0$. As the V and AV approach each other, the $\psi$ profile are extended. The time evolution of the system follows from (a) to (b).}
\label{fig4}
\end{figure}

 As can be seen in Fig.~\ref{fig4}, the vortex and the antivortex also present an elongated shape however, as the effective superconducting region is very wide, there is no channel liking the AD to the outer space. In the sequence, as the vortex and the antivortex approach each other, the region where superconductivity is practically suppressed - which is quite elongated in this system configuration - progressively overlap until the complete annihilation of the pair.

\begin{figure}
\includegraphics[width=0.9\columnwidth,height=1.4\linewidth]{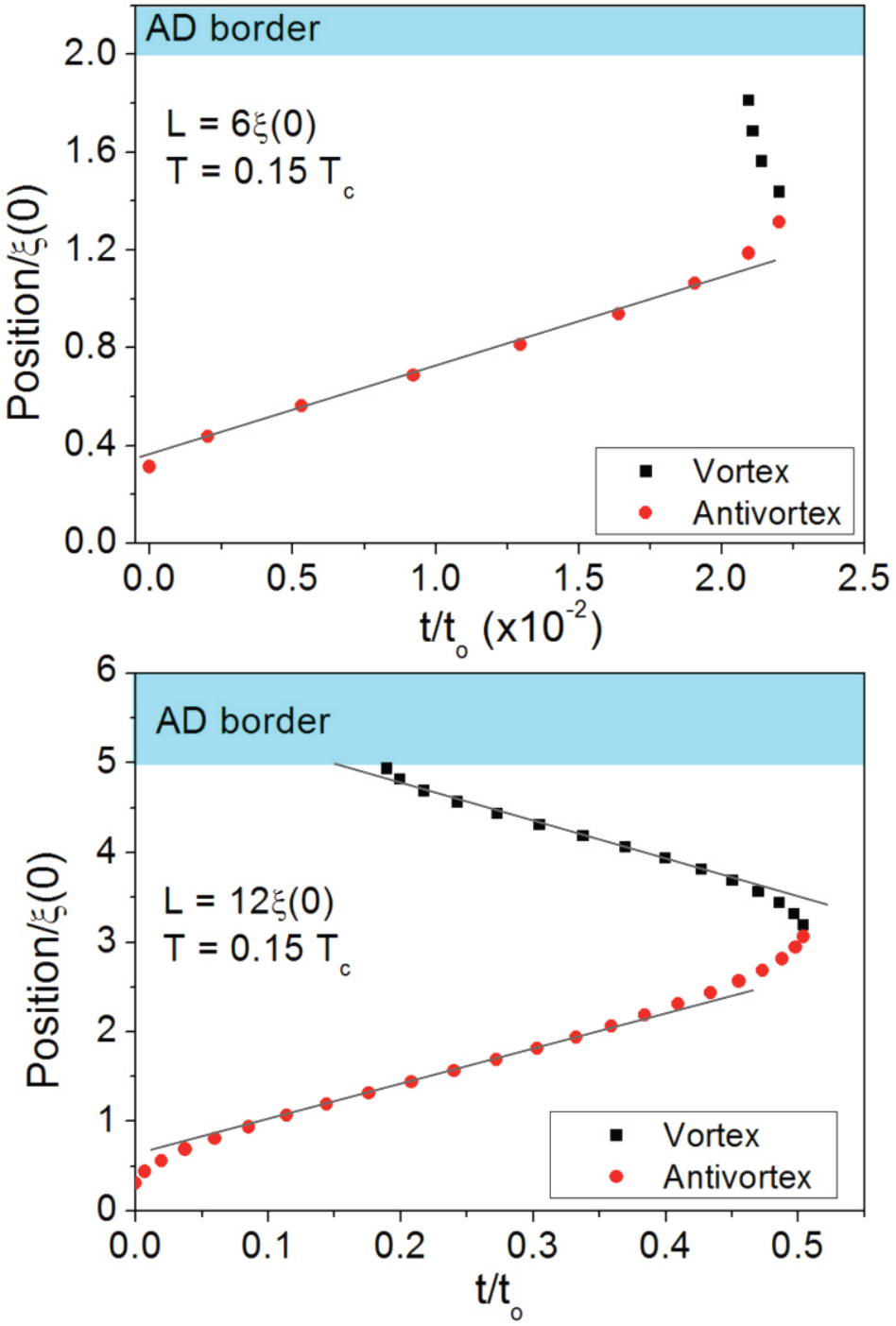}
\caption{(Color online) Normalized position versus normalized time for the annihilation motion of the V-AV pair in both studied systems. The straight lines indicate the motion of the V-AV pair with constant velocity.}
\label{fig5}
\end{figure}

 Figure~\ref{fig5} shows curves of normalized position versus normalized time for both studied systems. Note that the vortex (antivortex) presents an accelerated motion when leaving the AD (entering the system) due to the circulating currents. After that, a motion of constant velocity takes place as indicated in Fig.~\ref{fig5} by the straight lines and, as the vortex and the antivortex are approaching each other, motion is accelerated again until the pair annihilation.

\section{Conclusions}
\label{theend}

This work was carried out by using the TDGL equations in their discretized form to simulate superconducting samples with a central AD. It was shown that in the annihilation process, both the vortex and the antivortex acquire an elongated shape. In the smaller sample, a channel between the outer of the system and the AD is created. The motion of the vortex-antivortex pair was also analyzed and we observed that the movement is accelerated at the early and final stages of the annihilation process. However, a motion with constant velocity takes place between these two accelerated situations. A detailed study of such dynamics will be published soon elsewhere.

\textbf{Acknowledgements}

We thank the Brazilian Agencies CNPq, Fundunesp/PROPe and the S\~{a}o Paulo Research Foundation (FAPESP), grants 2013/11114-7 and  2007/08072-0, for financial support.

\bibliographystyle{model3a-num-names}

\begin{thebibliography}{00}

\bibitem{berdiyorov} G. R. Berdiyorov, A. D. Hern\'adez-Nieves,M. V. Milo\v{s}evi\'c, F. M. Peeters, D. Dom\'inguez, Phys. Rev. B \textbf{85}, 092502 (2012).
\bibitem{muller} A. M\"uller, M. V. Milo\v{s}evi\'c, S. E. C. Dale, M. A. Engbarth, S. J. Bending, Phys. Rev. Lett. \textbf{109}, 197003 (2012).
\bibitem{schweigert} V. A. Schweigert, F. M. Peeters, P. Singha Deo, Phys. Rev. Lett. \textbf{81}, 2783 (1998).
\bibitem{melnikov} A. S. Mel'nikov, I. M. Nefedov, D. A. Ryzhov, I. A. Shereshevskii, V. M. Vinokur, P. P. Vysheslavtsev,
Phys. Rev. B \textbf{65}, 140503(R) (2002).
\bibitem{mertelj} T. Mertelj, V. V. Kabanov, Phys. Rev. B \textbf{67}, 134527 (2003).
\bibitem{baelus} B. J. Baelus and F. M. Peeters, Phys. Rev. B \textbf{65}, 104515 (2002).
\bibitem{geim1} A. K. Geim, I. V. Grigorieva, S. V. Dubonos, J. G. S. Lok, J. C. Maan, A. E. Filippov, F. M. Peeters, Nature \textbf{390}, 259 (1997).
\bibitem{chibotaru}	L. F. Chibotaru, A. Ceulemans, V. Bruyndoncx, V. V. Moshchalkov, Nature \textbf{408}, 833 (2000).
%\bibitem{sardella1}	E. Sardella, P. N. Lisboa-Filho, C. C. de Souza Silva, L. R. E. Cabral, W. A. Ortiz, Phys. Rev. B \textbf{80}, 012506 (2009).
%\bibitem{grigorieva} I. V. Grigorieva, W. Escoffier, V. R. Misko, B. J. Baelus, F. M. Peeters, L. Y. Vinnikov, S. V. Dubonos, Phys. Rev. Lett. \textbf{99}, 147003 (2007).
\bibitem{moshchalkov} V. V. Moshchalkov, L. Gielen, C. Strunk, R. Jonckheere, X. Qiu, C. Van Haesendonck, Y. Bruynseraede, Nature \textbf{373}, 319 (1995).
\bibitem{sardella2} E. Sardella and E. H. Brandt, Supercond. Sci. Technol. \textbf{23}, 025015 (2010).
\bibitem{pascolati} M. C. V. Pascolati, E. Sardella, P. N. Lisboa-Filho, Physica C \textbf{470}, 206 (2010).
%\bibitem{yampolskii} S. V. Yampolskii, F. M. Peeters, Phys. Rev. B \textbf{62}, 9663 (2000).
%\bibitem{zhao} H. J. Zhao, V. R. Misko, F. M. Peeters, S. Dubonos, V. Oboznov, I. V. Grigorieva, EPL \textbf{83}, 17008 (2008).
\bibitem{benxu} Ben Xu, M. V. Milo\v{s}evi\'c,Shi-Hsin Lin, F. M. Peeters, B. Jank\'o, Phys. Rev. Lett. \textbf{107}, 057002 (2011).
\bibitem{cren} T. Cren, L. Serrier-Garcia, F. Debontridder, D. Roditchev, Phys. Rev. Lett. \textbf{107}, 097202 (2011).
%\bibitem{kanda} A. Kanda, B. J. Baelus, F. M. Peeters, K. Kadowaki, Y. Ootuka, Phys. Rev. Lett. \textbf{93}, 257002 (2004).
\bibitem{golubovic} D. S. Golubovic, M. V. Milo\v{s}evi\'c, F. M. Peeters, V. V. Moshchalkov, Phys. Rev. B \textbf{71}, 180502 (2005).
%\bibitem{kanda2} B. J. Baelus, A. Kanda, F. M. Peeters, Y. Ootuka, K.\ Kadowaki, Phys. Rev.B \textbf{71}, 140502 (2005).
\bibitem{milosevic} M. V. Milo\v{s}evi\'c, A. Kanda, S. Hatsumi, F. M. Peeters, Y. Ootuka, Phys.Rev.Lett. \textbf{103}, 217003 (2009).
\bibitem{yu} A. Yu. Aladyshkin, I. M. Nefedov, A. S. Aladyshkina, I. A. Shereshevskii, Physica C \textbf{479}, 98 (2012).
\bibitem{geim3}	A. K. Geim, S. V. Dubonos, I. V. Grigorieva, K. S. Novoselov, F. M. Peeters, V. A. Schweigert, Nature \textbf{407}, 55 (2000).
\bibitem{buzdin} A. I. Buzdin, J. P. Brison, Phys. Lett. A \textbf{196}, 267 (1994).
\bibitem{palacios} J. J. Palacios, Phys. Rev. Lett. \textbf{84}, 1796 (2000).
%\bibitem{baelus2} B. J. Baelus, F. M. Peeters, V. A. Schweiger, Phys. Rev. B \textbf{63}, 144517 (2001).
%\bibitem{cabral} L. R. E. Cabral, B. J. Baelus, F. M. Peeters, Phys. Rev. B \textbf{70}, 144523 (2004).
\bibitem{zad2} R. Zadorosny, E. Sardella, A. L. Malvezzi, P. N. Lisboa Filho, and W. A. Ortiz, Phys. Rev. B \textbf{85}, 214511 (2012).
%\bibitem{misko} V. R. Misko, B. Xu, F. M. Peeters, Phys. Rev. B \textbf{76}, 024516 (2007).
%\bibitem{sardella3} E. Sardella, P. N. Lisboa-Filho, A. L. Malvezzi, Phys. Rev. B \textbf{77}, 104508 (2008).
\bibitem{misko2} V.R. Misko, B. Xu, F.M. Peeters, Physica C \textbf{468}, 726(2008).
%\bibitem{zhao2} H. J. Zhao, V. R. Misko, F. M. Peeters, V. Oboznov, S. V. Dubonos, I. V. Grigorieva, Phys. Rev. B \textbf{78}, 104517 (2008).
\bibitem{zad1} R. Zadorosny, E. Sardella, A. L. Malvezzi, P. N. Lisboa Filho, and W. A. Ortiz, Physca C \textbf{479}, 154 (2012).
\bibitem{connolly} M. R. Connolly, M. V. Milo\v{s}evi\'c, S. J. Bending, J. R. Clem, T. Tamegai, EPL \textbf {85}, 17008 (2009).
\bibitem{sardella4} E. Sardella, A. L. Malvezzi, P. N. Lisboa-Filho, W. A. Ortiz, Phys. Rev. B \textbf{74},014512 (2006).
\bibitem{geurts} R. Geurts, M. V. Milo\v{s}evi\'c, F. M. Peeters, Phys. Rev. B \textbf{81}, 214514 (2010).
\bibitem{geurts2} M. V. Milo\v{s}evi\'c, R. Geurts, Physica C \textbf{470}, 791 (2010).
\bibitem{gropp} W. D. Gropp, H. G. Kaper, G. K. Leaf, D. M. Levine, M. Palumbo, and V. M. Vinokur,J. Comput. Phys. \textbf{123}, 254 (1996).


 \end{thebibliography}

\end{document}